\documentclass[twocolumn,prl,superscriptaddress]{revtex4}
\usepackage{epsf}
\newcommand{\beq}{\begin{equation}}
\newcommand{\eeq}{\end{equation}}
\newcommand{\be}{\begin{eqnarray}}
\newcommand{\ee}{\end{eqnarray}}
\begin{document}
\title{Polarized antiquark flavor asymmetry: \\
Pauli blocking vs.\ the pion cloud}
\author{R.~J.~Fries}
\affiliation{Institut f\"ur Theoretische Physik,
Universit\"at Regensburg, D--93053 Regensburg, Germany}
\affiliation{Physics Department, Duke University, 
P.O. Box 90305, Durham, NC 27708}
\author{A.~Sch\"afer} 
\affiliation{Institut f\"ur Theoretische Physik,
Universit\"at Regensburg, D--93053 Regensburg, Germany}
\author{C.~Weiss}
\affiliation{Institut f\"ur Theoretische Physik,
Universit\"at Regensburg, D--93053 Regensburg, Germany}
\begin{abstract}
The flavor asymmetry of the unpolarized antiquark distributions in the 
proton, $\bar d(x) - \bar u(x) > 0$, can qualitatively be explained either by
Pauli blocking by the valence quarks, or as an effect of the pion cloud of 
the nucleon. In contrast, predictions for the polarized 
asymmetry $\Delta \bar u(x) - \Delta \bar d(x)$ based on $\rho$ meson 
contributions disagree even in sign with the Pauli blocking picture. 
We show that in the meson cloud picture
a large positive $\Delta \bar u(x) - \Delta \bar d(x)$ 
is obtained from $\pi N$--$\sigma N$ interference--type contributions, 
as suggested by chiral symmetry. This effect restores the equivalence 
of the ``quark'' and ``meson'' descriptions also in the polarized case.
\end{abstract}
\maketitle
That the low--energy structure of the nucleon can be described equally 
well in terms of quark or meson degrees of freedom
has been one of the fundamental beliefs of modern hadronic physics.
While often one description is far more efficient than the other,
there is a basic conviction that both should give equivalent
results when carried on to higher accuracy, which, unfortunately,
often turns out to be impossible in practice.
\par
Particularly interesting properties in this respect
are the parton (quark-- and antiquark) distributions in the nucleon.
Although measured in deep--inelastic 
scattering at large momentum transfers, these
are low--energy characteristics of the nucleon,
whose origin can be understood on grounds of the same effective
dynamics which gives rise to the hadronic characteristics of the
nucleon such as form factors, magnetic moments, {\it etc.}
\par
It is now well established that the antiquark distributions
in the proton are not flavor symmetric: $\bar d (x) > \bar u(x)$.
Deep--inelastic lepton scattering has convincingly demonstrated the
violation of the so--called Gottfried sum rule \cite{Kumano:1997cy}, 
and the E866 Drell--Yan pair production data \cite{Hawker:1998ty} as well 
as the HERMES results on semi-inclusive deep--inelastic 
scattering \cite{Ackerstaff:1998sr}
allow to map even the $x$--dependence of the asymmetry.
The origin of this asymmetry can qualitatively be explained
in either a quark or a meson picture. In the quark picture
it can be attributed to the ``Pauli blocking'' 
effect \cite{Field:ve,Schreiber:tc,Melnitchouk:1998rv}. For instance, in the 
bag model, where the valence quarks are bound by a scalar field, the Dirac 
vacuum inside the proton differs from the free one, corresponding to the
presence of a non-perturbative ``sea'' of quark--antiquark pairs.
Since the wave function of a localized valence quark in the proton rest 
frame has components corresponding to antiquarks in the 
infinite--momentum frame, the valence quarks ``block'' quark--antiquark 
pairs of the same flavor, leading to an excess of $\bar d (x)$ over 
$\bar u (x)$ \cite{Schreiber:tc}. The mesonic picture attributes the antiquark
flavor asymmetry to the contribution of the ``pion cloud'' of the 
proton to deep--inelastic scattering (Sullivan mechanism) 
\cite{Sullivan:1971kd,Thomas:1983fh}. The asymmetry arises because
fluctuations $p \rightarrow n \pi^+$ are more likely than 
$p \rightarrow \Delta^{++} \pi^-$ due to the larger mass of the 
$\Delta$ resonance, which implies a larger number of $\pi^+$
than $\pi^-$ in the proton's cloud. It needs to be stressed that both 
explanations are of qualitative nature; the difficulties encountered
when trying to turn them into serious dynamical models 
have been discussed in the literature, see {\it e.g.}\ 
Refs.\cite{Steffens:1996bc,Koepf:1995yh}. Nevertheless, the fact that
the two pictures give compatible results for the 
sign and order--of--magnitude of $\bar d(x) - \bar u(x)$
has been registered as a remarkable instance of equivalence of
a quark and a meson description.
\par
Recently the polarized antiquark flavor asymmetry, 
$\Delta\bar u (x) - \Delta\bar d (x)$, has become a focus of attention. 
It is expected that this asymmetry will be measured with good accuracy
in polarized semi-inclusive particle production at the HERMES experiment, 
and, in particular, in future polarized Drell--Yan pair or $W^\pm$ production 
experiments at 
RHIC \cite{McGaughey:1999mq,Soffer:1997fe,Dressler:1999zv,Kumano:1999bt}. 
The published semi-inclusive data from HERMES \cite{Ackerstaff:1999ey} 
and SMC \cite{Adeva:1995yi} do not yet allow for significant 
conclusions \cite{Dressler:1999zg}; improved data from HERMES are
expected to be released soon. On the theoretical side,
interest was caused by an estimate within the chiral quark--soliton model 
of the nucleon, based on the large--$N_c$ limit of QCD, which 
suggests a surprisingly large positive
$\Delta \bar u(x) - \Delta \bar d(x)$, larger than the unpolarized 
asymmetry, $\bar d(x) - \bar u(x)$ \cite{Diakonov:1996sr}. 
\par
It is natural to ask what the two standard explanations for the
unpolarized asymmetry predict for the polarized case. The Pauli blocking 
picture implies that valence quarks ``block'' antiquarks of the same
flavor but with opposite spin, which would give
$\Delta \bar u(x) - \Delta \bar d(x) > 0$ \cite{Schreiber:tc,Cao:2001nu}. 
Gl\"uck and Reya \cite{Gluck:2000ch} have suggested a phenomenological 
parametrization based on the ansatz $\Delta \bar u(x) / \Delta \bar d(x)
= \Delta d(x) / \Delta u(x)$, which qualitatively expresses this
idea. The resulting asymmetry at a scale of 
$\mu^2 = 1\,{\rm GeV}^2$, as obtained with the 
AAC parametrization \cite{Goto:1999by}
of the polarized valence and flavor--singlet sea quark 
distributions, is shown in Fig.~\ref{fig_models} (dotted line). 
It should be stressed, however, that 
the simple Pauli blocking argument can predict neither the magnitude nor 
the $x$--dependence of the polarized asymmetry. Nevertheless, it is
natural to assume that in such a picture the polarized asymmetry should 
be of the same order of magnitude as the unpolarized one \cite{Cao:2001nu}.
\par
\begin{figure}[ht]
\begin{center}
\setlength{\epsfxsize}{8.4cm}
\setlength{\epsfysize}{5.94cm}
\epsffile{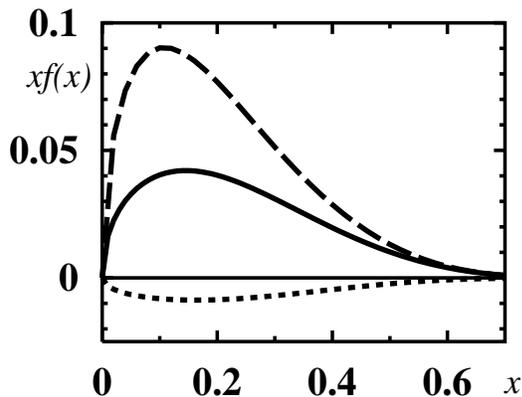}
\end{center}
\caption[]{Various contributions to the antiquark flavor asymmetry 
in the proton (unpolarized and polarized) in the meson cloud model
(scale $\mu^2 = 1\,{\rm GeV}^2$). 
{\it Dashed line:} $x [ \bar d (x) - \bar u(x) ]$, $\pi N$
contributions (Sullivan mechanism). 
{\it Dotted line:} $x [ \Delta \bar u (x) - 
\Delta \bar d(x)]$, $\rho N$ contribution \cite{Fries:1998at}.
{\it Solid line:} $x [ \Delta \bar u (x) - \Delta \bar d(x)]$,
$\pi N$--$\sigma N$ interference contribution.}
\label{fig_cloud}
\end{figure}
\begin{figure}[ht]
\begin{center}
\setlength{\epsfxsize}{8.4cm}
\setlength{\epsfysize}{5.94cm}
\epsffile{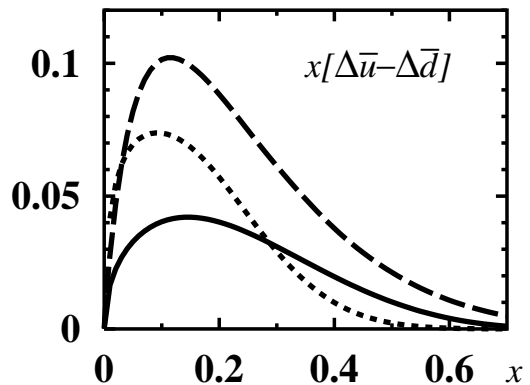}
\end{center}
\caption[]{Comparison of model results for the
polarized flavor asymmetry 
$x [\Delta \bar u (x) - \Delta \bar d(x) ]$
in the proton ($\mu^2 = 1\,{\rm GeV}^2$).
{\it Dotted line:} Pauli blocking ansatz of Ref.\cite{Gluck:2000ch}.
{\it Dashed line:} Chiral quark--soliton model \cite{Diakonov:1996sr}.
{\it Solid line:} $\pi N$--$\sigma N$ interference contribution 
in the meson cloud model ({\it cf.}\ Fig.~\ref{fig_cloud}).}
\label{fig_models}
\end{figure}

In the meson cloud picture, the $\pi N$ contribution 
(Sullivan mechanism) gives zero polarized flavor asymmetry. 
The inclusion of $\rho N$ contributions \cite{Fries:1998at} leads to a non-zero 
$\Delta\bar u (x) - \Delta\bar d (x)$, 
which, however, is an order of magnitude smaller than the unpolarized one,
and has opposite sign to what is expected from Pauli
blocking, see Fig.~\ref{fig_cloud} (dotted line). 
Ref.\cite{Boreskov:1998hp} studied $\rho N$--$\pi N$
interference contributions relevant at small $x$, see also
Ref.\cite{Cao:2001nu}, the sign again opposite to Pauli blocking.
The estimate of Ref.\cite{Fries:1998at} was refined by including also 
higher--twist components of the $\rho$ meson structure 
function \cite{Kumano:2001cu}, which, however, do not change the order 
of magnitude of the result. One may thus wonder whether the equivalence 
of the quark and meson descriptions, which was observed in the unpolarized 
asymmetry, fails in the polarized case.
\par
Here we want to demonstrate that a negative polarized antiquark flavor 
asymmetry is by no means a necessary consequence of the meson cloud picture.
In fact, a sizable positive $\Delta\bar u (x) - \Delta\bar d (x)$
is naturally obtained from $\pi N$--$\sigma N$ ``interference type'' 
contributions to the nucleon parton distributions. The possibility of
such contributions was first pointed out in 
Ref.\cite{Dressler:1999zg}. Here $\sigma$ means the 
scalar--isoscalar meson, which appears as the chiral partner of the pion, 
and which mediates the intermediate--range $NN$ interaction in the 
meson exchange parametrization of Ref.\cite{Machleidt:hj}
(Bonn potential). The problem of the dynamical nature of the $\sigma$ 
meson --- whether it should be regarded as an effective description
of a $\pi\pi$ resonance --- has been discussed extensively in the
literature and shall not concern us here.
\par
\begin{figure}[ht]
\begin{center}
\setlength{\epsfxsize}{8.4cm}
\setlength{\epsfysize}{3.5cm}
\epsffile{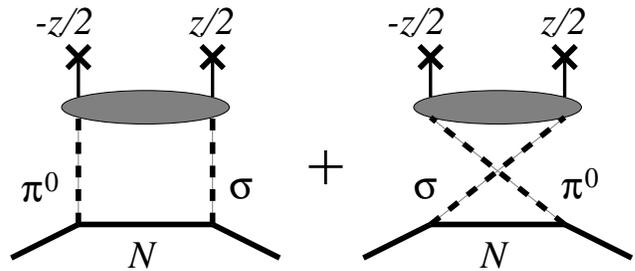}
\end{center}
\caption[]{$\pi N$--$\sigma N$ ``interference type'' graphs contributing
to $\Delta \bar u (x) - \Delta \bar d(x)$ in the proton. The crosses
denote the positions of the quark fields in the QCD twist--2 operator,
$z/2$ and $-z/2$, {\it cf.}\ Eq.(\ref{bosonized_operator}).}
\label{fig_graphs}
\end{figure}

To illustrate our point we have computed the $\pi N$--$\sigma N$ interference 
contributions to $\Delta\bar u (x) - \Delta\bar d (x)$ in the proton
in a linear sigma model with elementary $\pi$ and $\sigma$ fields
coupled to the nucleon. The isovector polarized quark-- and antiquark 
distributions in the proton are defined by the matrix element of the 
twist--2 axial vector light--ray operator:
\be
&&
\int\frac{dz^-}{2\pi} e^{\pm i x p^+ z^-} \langle p |
\bar\psi (-z/2) \gamma^+ \gamma_5 \tau^3 \psi (z/2) 
| p \rangle_{z^+, {\bf z}_\perp =0} 
\nonumber \\
&=& \bar U \gamma^+ \gamma_5 U \times
\left\{
\begin{array}{lcl}
\left[ \Delta u (x) - \Delta d (x) \right] 
\\[1ex]
\left[ \Delta \bar u (x) - \Delta \bar d (x) \right]
\end{array} .
\right.
\ee
Here $0 < x < 1$, and $z^{\pm} = (z^0 \pm z^3)/\sqrt{2}$ and ${\bf z}_\perp$ 
are the usual light--like coordinates, $\tau^3$ the isospin Pauli matrix,
and $\bar U, U$ the proton spinors. We consider the contribution to the matrix 
element from the ``interference type'' graphs of Fig.~\ref{fig_graphs}.
The blob in the upper parts of the graphs denotes the ``bosonized'' version 
of the isovector axial vector twist--2 operator, {\it i.e.}, the operator 
expressed in terms of the $\pi$ and $\sigma$ fields
of our effective low--energy model. We suppose here that the QCD operator 
is normalized at a scale of $\sim 1 \, {\rm GeV}$, up 
to which the effective model is assumed to be valid. On general grounds 
the matching of the QCD operator to an operator in the effective model 
must be of the form
\be
&& \bar\psi (-z/2) \gamma^+ \gamma_5 \tau^a \psi (z/2) |_{z^2=0}
\nonumber \\
&\rightarrow&
\int_{-1}^1 dy \; g_{\pi\sigma} (y) \;
\sigma (-y z/2) \stackrel{\leftrightarrow}{\partial}{}^{\!\! +}
\pi^a (yz/2)|_{z^2=0} ,
\label{bosonized_operator}
\ee
up to terms of higher orders in derivatives of the fields, 
which we shall neglect. Here $g_{\pi\sigma} (y)$ is a scalar function, 
which we refer to as the ``$\pi$--$\sigma$ transition parton density''. 
The expansion of Eq.(\ref{bosonized_operator}) in powers of the light--like 
distance, $z$, implies that the local twist--2 spin--$n$ operator is mapped 
onto the local twist--2 spin--$n$ operator built from the $\pi$
and $\sigma$ fields, with the coefficient given by the $n$'th moment
of $g_{\pi\sigma}$. Time reversal invariance requires 
$g_{\pi\sigma}(y) = g_{\pi\sigma}(-y)$. The normalization of the
function follows from considering the limit $z \rightarrow 0$, in which
the R.H.S.\ of Eq.(\ref{bosonized_operator}) must reduce to the
isovector axial current operator in the $\pi$ and $\sigma$ fields,
$\sigma (0) \stackrel{\leftrightarrow}{\partial}{}^{\!\! \mu} \pi^a (0)$,
whose form is completely determined by chiral symmetry. This requires
\beq
\int_{-1}^1 dx \; g_{\pi\sigma} (y) \;\; = \;\; 2.
\eeq
In order to constrain the $y$--dependence of $g_{\pi\sigma}$
we note that a global chiral rotation transforms the axial vector 
operators of Eq.(\ref{bosonized_operator}) into the 
corresponding vector operators, whose matrix element between 
pion states defines the valence quark distribution in the pion,
$v_\pi (y)$. Thus, in our approximation we can identify
\beq
g_{\pi\sigma} (y) \;\; = \;\; \frac{1}{2} v_\pi (|y|) .
\eeq
In our estimate we use the parametrization of Ref.\cite{Gluck:1999xe} 
for $v_\pi (y)$, obtained from fitting $\pi N$ Drell--Yan data.
\par
The contribution of the 
two graphs of Fig.~\ref{fig_graphs}
to the polarized flavor asymmetry can be put in the form
\be
\Delta \bar u(x) - \Delta \bar d(x) 
&=& \int_x^1 \frac{dy}{y} \; g_{\pi\sigma} (y) \; 
W_{\pi\sigma} \left( \frac{x}{y} \right)
\ee
where $W_{\pi\sigma} (x/y)$ denotes the correlation
function of the $\pi$-- and $\sigma$--fields in the nucleon
depending on the $+$ component of the fields' momenta
($v \equiv x/y$)
\be
W_{\pi\sigma}(v) &=& \frac{g_{\pi NN} g_{\sigma NN}}{4\pi} 
\int\frac{d^2 k_\perp}{(2\pi )^2}
\nonumber \\
&& \times 
\frac{x [{\bf k}_\perp^2 + v (2 - v) M_N^2 ]}
{\left[ {\bf k}_\perp^2 + v^2 M_N^2 + (1 - v) M_\pi^2 \right]}
\nonumber \\
&& \times \frac{1}{\left[ {\bf k}_\perp^2 + v^2 M_N^2 
+ (1 - v) M_\sigma^2 \right]} .
\label{W}
\ee
This function
plays a role analogous to the ``number of pions with momentum
fraction $v$'' in the usual $\pi N$ contribution to the unpolarized 
asymmetry \cite{Thomas:1983fh}. The integral over the transverse
momentum ${\bf k}_\perp$ contains a would--be logarithmic divergence
which is regularized by cutoffs associated with the
$\pi N$ and $\sigma N$ vertices, not indicated in Eq.(\ref{W}).
For a numerical estimate we use coupling constants
$g_{\pi NN} = 13.5, g_{\sigma NN} = 14.6$,  
$M_\sigma = 0.72\, {\rm GeV}$, and exponential
cutoffs with $\Lambda_\pi = 1.1\, {\rm GeV}$
and $\Lambda_\sigma = 1.6\, {\rm GeV}$ \cite{Machleidt:hj}.
The result for $\Delta \bar u(x) - \Delta \bar d(x)$ is shown by the
solid line in Fig.~\ref{fig_cloud}. The dashed line in the same figure
shows the $\pi N$ contribution to the unpolarized asymmetry,
$\bar d(x) - \bar u(x)$, evaluated with the same parameters.
One sees that the polarized asymmetry incurred from $\pi N$--$\sigma N$
interference is positive, and of the same order of magnitude
as the unpolarized one. (In Fig.~\ref{fig_cloud}, for the sake of 
comparison, we show $\bar d(x) - \bar u(x)$ as generated by $\pi N$
contributions only; it is known that the inclusion of intermediate 
$\Delta$ states reduces this value by almost 50\% \cite{Thomas:1983fh}.)
\par
In Fig.~\ref{fig_models} we compare the $\pi N$--$\sigma N$
interference contribution to $\Delta \bar u(x) - \Delta \bar d(x)$
(solid line) with the asymmetry obtained with the phenomenological
Pauli--blocking ansatz of Ref.\cite{Gluck:2000ch}. One sees that both
suggest a sizable positive flavor asymmetry 
$\Delta \bar u(x) - \Delta \bar d(x)$. In this sense, one may say 
that the same qualitative equivalence of the ``quark'' and ``meson'' 
descriptions holds as in the case of unpolarized asymmetry, 
$\bar d (x) - \bar u(x)$.
\par
We stress that our point here is entirely qualitative, concerning only
the sign and order--of--magnitude of the asymmetry. Neither the
Pauli--blocking ansatz of Ref.~\cite{Gluck:2000ch} nor the $\pi N$--$\sigma N$
contribution in the meson cloud model can claim to give a quantitative
description of the $x$--dependence of the asymmetry. (We also refrain from
quoting any error estimate for the meson cloud model.) Nevertheless,
given the disagreement even in sign of the previous $\rho$ meson cloud 
estimates with the Pauli blocking picture we feel that the agreement
at the present level is remarkable. Note also that in the near future 
experiments will be able to determine little more but the sign and 
order--of--magnitude of $\Delta \bar u(x) - \Delta \bar d(x)$.
\par
Also in Fig.~\ref{fig_models}, we compare the estimates from the
$\pi N$--$\sigma N$ contribution in the meson cloud model and the 
Pauli blocking ansatz of Ref.\cite{Gluck:2000ch} with the result of the 
chiral quark--soliton model (ChQSM), which was the
first to predict a large positive flavor
asymmetry $\Delta \bar u (x) - \Delta \bar d(x)$ \cite{Diakonov:1996sr}. 
This comparison
is interesting also from a conceptual point of view. In the
ChQSM, motivated by the large--$N_c$ limit of QCD \cite{Witten:1979kh}, 
the nucleon is described by a classical pion field, in which quarks 
move in single--particle orbits. The quark spectrum includes a 
bound--state level in addition to the polarized negative and positive
Dirac continua \cite{Diakonov:1987ty}. In a sense, this model contains the 
physical essence of both the ``Pauli blocking'' and the ``meson cloud'' 
picture, uniting both of them in a consistent framework. The contribution
of the bound--state level of quarks to $\Delta \bar u (x) - \Delta \bar d(x)$
is positive, in agreement with the ``Pauli blocking'' argument. The
contribution to $\Delta \bar u (x) - \Delta \bar d(x)$
from the Dirac sea of quarks can be computed approximately in
an expansion in gradients of the classical pion field polarizing
the vacuum; the result is given as a spatial integral of the
the isovector--pseudoscalar and scalar--isoscalar combinations
of the classical field, reminiscent in quantum numbers of the
$\pi N$--$\sigma N$ interference contribution in the meson cloud 
model \cite{Dressler:1999zg}. Thus, the semiclassical description of the 
nucleon at large $N_c$ reproduces the physics of ``meson cloud''
contributions to the nucleon parton distributions without appealing
to the notion of individual meson exchange graphs. In this way it avoids 
the conceptual problems of the meson cloud model related to the neglection 
of multiple exchanges and the large virtuality of the exchanged 
mesons (see Ref.\cite{Koepf:1995yh} for a critical discussion).
\par
To summarize, we have argued that the ``Pauli blocking'' and the 
``meson cloud'' scenario are both consistent with a positive 
polarized antiquark flavor asymmetry, $\Delta \bar u (x) - \Delta \bar d(x)$,
of comparable magnitude as the unpolarized one, 
$\bar d(x) - \bar u(x)$. The key to this equivalence has been
the inclusion of $\pi N$--$\sigma N$ ``interference type'' contributions
in the meson cloud picture, whose importance is suggested by chiral symmetry. 
Our qualitative arguments explain the large value of 
$\Delta \bar u (x) - \Delta \bar d(x)$ predicted by the
chiral quark--soliton model. This should be good news
for experiments aimed at extracting $\Delta \bar u (x) - \Delta \bar d(x)$,
both from semi-inclusive deep--inelastic scattering and 
Drell--Yan / $W^\pm$ production.
\par
We are grateful to M.~V.~Polyakov for many helpful suggestions,
and to A.~W.~Thomas and W.~Melnitchouk for useful discussions.
R.~J.~F. is supported by the Alexander von Humboldt Foundation
(Feodor Lynen Fellowship), C.~W.~by DFG (Heisenberg Fellowship). 
This work has been supported by DFG and BMBF.
\end{document}